\documentclass[fleqn,a4paper,12pt]{article}
\usepackage{amssymb}
\usepackage{amsmath}
\usepackage[dvips]{graphicx}
\usepackage{placeins}
\usepackage{lscape}
\usepackage{a4}
\usepackage{epsfig}

\newcommand{\beq}{\begin{equation}} \newcommand{\eeq}{\end{equation}}
\newcommand{\bea}{\begin{array}} \newcommand{\eea}{\end{array}}
\newcommand{\ri}{{\mathrm i}}

\catcode`@=11 \long
\def\@caption#1[#2]#3{\par\addcontentsline{\csname
ext@#1\endcsname}{#1} {\protect\numberline{\csname
the#1\endcsname}{\ignorespaces #2}} \begingroup \small
\@parboxrestore \@makecaption{\csname fnum@#1\endcsname}
{\ignorespaces #3}\par \endgroup} \catcode`@=12

\newcommand{\so}{\sigma_1}
\newcommand{\sd}{\sigma_2}
\newcommand{\st}{\sigma_3}

\newcommand{\p}{\partial}

\newcommand{\R}{\mathbb{R}}

\DeclareMathOperator{\sech}{sech}
\DeclareMathOperator{\csch}{csch}

\begin{document}
\allowdisplaybreaks
 \begin{titlepage} \vskip 2cm

\begin{center} {\Large\bf Matrix superpotentials} \footnote{E-mail:
{\tt nikitin@imath.kiev.ua},\ \ {\tt yuri.karadzhov@gmail.com}  }
\vskip 3cm {\bf {A.G. Nikitin$^a$ and Yuri Karadzhov$^a$} \vskip 5pt
{\sl $^a$Institute of Mathematics, National Academy of Sciences of
Ukraine,\\ 3 Tereshchenkivs'ka Street, Kyiv-4, Ukraine, 01601\\}}
\end{center}
\vskip .5cm \rm
\begin{abstract} We present a collection of matrix valued shape invariant
potentials which give rise to new exactly solvable problems of SUSY
quantum mechanics. It includes all irreducible matrix
superpotentials of the generic form $W=kQ+\frac1k R+P$ where $k$ is
a variable parameter, $Q$ is the unit matrix multiplied by a real
valued function of independent variable $x$, and $P$, $R$ are
hermitian matrices depending on $x$. In particular we recover the
Pron'ko-Stroganov "matrix Coulomb potential" and all known scalar
shape invariant potentials of SUSY quantum mechanics.

In addition, five new shape invariant potentials are presented.
Three of them admit a dual shape invariance, i.e., the related
hamiltonians can be factorized using two non-equivalent
superpotentials.  We find discrete spectrum and eigenvectors for the
corresponding Schr\"odinger equations and prove that these
eigenvectors  are normalizable.

\end{abstract}

\end{titlepage}

\section{Introduction\label{intro}}

  Invented by E. Witten \cite{Witten} as a toy model supersymmetric quantum
  mechanics (SSQM)
  became a fundamental field including many interesting external and internal problems.
In particular the SSQM  presents powerful tools for explicit
solution of quantum mechanical problems using the shape invariance
approach \cite{Gen}. Unfortunately, the number of problems
satisfying the shape invariance condition is rather restricted.
However, such problems include practically all cases when the
related Schr\"odinger equation is exactly solvable and has an
explicitly presentable potential. Well known exceptions are exactly
solvable Schr\"odinger equations with  Natanzon potentials
\cite{natan} which are formulated in terms of implicit functions.
The list of shape invariant potentials depending on one variable can
be found in \cite{Khare}.

An interesting example of QM problem which admits a shape invariant
supersymmetric formulation was discovered by Pron'ko and Stroganov
\cite{Pron} who studied a motion of a neutral non-relativistic
fermion which interacts anomalously with the magnetic field
generated by a thin current carrying wire.

The supersymmetric approach to the Pron'ko-Stroganov (PS) problem
was first applied in paper \cite{Vor} with using the momentum
representation. In paper \cite{Gol} this problem was solved using its
 shape invariance in the coordinate representation. Recently a
 relativistic generalization
of the PS problem was proposed \cite{ninni} which can also be
integrated using its supersymmetry with shape invariance.

 The specificity
 of the PS problem is that it is
 formulated
 using  a {\it matrix superpotential} while in the standard SSQM
  the superpotential is simply a scalar function. Matrix
  superpotentials themselves were discussed in many papers, see, e.g.,
  \cite{Andr}, \cite{Andri}, \cite{Fu}, \cite{Rodr} but this
  discussion was actually
  reduced to analysis of particular examples. In papers \cite{tkach}
  such superpotentials were used for analysis of motion of a spin
  $\frac12$ particle in superposed magnetic and scalar fields.
  In paper \cite{Fu} a certain class of such superpotentials was described
  which however was {\it ad hoc} restricted to $2\times2$ matrices which
  depend linearly on the variable parameter. Thus, in contrast to the case
  of scalar superpotentials, the class of known matrix potentials includes
  only few examples which are important but rather particular, while the
  remaining part of this class is still
  "terra incognita".  It seems to be interesting
  to extend our knowledge of these potentials since this way it is
  possible to find new systems of Schr\"odinger equations which are
  exactly integrable.

  In the present paper a certain class of matrix valued superpotentials is
  described which includes the superpotential for the
  PS problem as a particular case. Moreover, we do not make {\it a
  priori }
  suppositions about the dimension of the involved matrices but restrict
  ourselves to linear and inverse dependence of the superpotentials on
  variable parameters. Trough our approach however the problem of
  classification of indecomposable matrix potentials which are shape invariant appears to be
  completely solvable. We present a classification of such
  potentials and discuss the corresponding
  exactly solvable problems for coupled systems of Schr\"odinger
  equations. In particular the discrete energy spectra and exact solutions for
  these models are found and normalizability of the ground and exited states
  is proven. Solutions corresponding to continuous spectra  are not
  considered here.

 Three out of five found hamiltonians admit  alternative factorizations
 with using different superpotentials. The corresponding potentials are
 shape invariant w.r.t. shifts of two different parameters.
 Such dual shape invariance results in existence of two alternative
 spectra branches. Moreover, for some values of free parameters both
 these branches can be realized.

\section{Superpotential for PS problem\label{PS}}

The PS problem was discussed in numerous papers, see, e.g.,
\cite{Pron}-\cite{Gol}. Thus we will not present its physical
motivations and  calculation details  but start with the
corresponding equation for radial functions \cite{Gol}
\begin{gather}
\label{eq}
\hat{H}_\kappa\psi=E_\kappa\psi
\end{gather}
where $\hat{H}_\kappa$ is a Hamiltonian with a matrix potential,
$E_\kappa$ and $\psi$ are  its eigenvalue and eigenfunction
correspondingly, moreover, $\psi$ is a two-component spinor. Up to
normalization of the radial variable $x$ the Hamiltonian $\hat{H}_\kappa$
can be represented as
\begin{gather}
\label{Ham}
\hat{H}_\kappa=-\frac{\partial^2}{\partial x^2}+\kappa(\kappa-\st)
\frac{1}{x^2}+\so\frac{1} {{x}}
\end{gather}
where $\so$
and $\st$ are Pauli matrices and $\kappa$ is a natural number. In
addition, solutions of equation (\ref{eq}) must be normalizable and
vanish at $x=0$.

Hamiltonian $\hat{H}_\kappa$ can be factorized as
\begin{equation}
\label{s3}
\hat{H}_\kappa=a_\kappa^+a_\kappa^-+c_\kappa
\end{equation}
where
\[
a_\kappa^-=\frac{\partial}{\partial x}+W_\kappa,\  \ a_\kappa^+=-
\frac{\partial}{\partial x}+W_\kappa,\ \
c_\kappa=-\frac{1}{(2\kappa+1)^2}
\]
and $W$ is a {\it matrix
superpotential}
\begin{equation}
\label{s4}
W_\kappa=\frac{1}{2x}\st-\frac{1}{2\kappa+1}\so-\frac{2\kappa+1}{2x}.
\end{equation}

Another nice property of Hamiltonian $\hat{H}_\kappa$ is that its
superpartner $\hat{H}^+_\kappa$ is equal to $\hat{H}_{\kappa+1}$,
namely
\[
\hat{H}_\kappa^+=a_\kappa^-a_\kappa^++c_\kappa= -\frac{\partial^2}{\partial x^2}+(\kappa+1)(\kappa+1-\st)\frac{1}{x^2}+\so\frac1x=\hat{H}_{\kappa+1}
\]
Thus equation (\ref{eq}) admits
supersymmetry with shape invariance and can be solved using the
standard technique of SSQM \cite{Khare}.

\section{Generic matrix shape invariant potentials\label{matrixproblem}}

Following a natural desire to find other matrix potentials which are
form invariant we consider equation (\ref{eq}) with
\begin{equation}
\label{hamiltonian} H_k=-\frac{\partial^2}{\partial x^2}+V_k(x),
\end{equation}
where $V_k(x)$ is an $n\times n$ dimensional matrix potential
depending on variable $x$ and parameter $k$.

Suppose that the Hamiltonian accepts factorization (\ref{s3}) where
$W_k(x)$ is a superpotential. Our goal is to find such
superpotentials which generate form invariant potentials $V_k(x)$.

Assume $W_k(x)$ is Hermitian. Then the corresponding
potential $V_k(x)$ and its superpartner $V^+_k(x)$, i.e.,
\begin{gather}
\label{pfactorization} V_k(x)=-\frac{\partial W_k}{\partial
x}+W_k^2\ \texttt{ and }\ V_k^+(x)=\frac{\partial W_k}{\partial
x}+W_k^2
\end{gather}
are Hermitian too.

Suppose also that the Hamiltonian be shape invariant, i.e.,
\begin{equation}
\label{hforminvariance} H_k^+=H_{k+\alpha}+C_k,
\end{equation}
thus
$
 V_k^+=V_{k+\alpha}+C_k
$ or
\begin{gather}\label{SI}W_k^2+W'_k=W_{k+\alpha}^2-
W'_{k+\alpha}+C_k\end{gather}
 were $C_k$ and $\alpha$ are constants.

 Let us state the problem of classification of shape invariant
 superpotentials, i.e., $n\times n$ matrices whose elements are functions
 of $x, k$  satisfying conditions (\ref{SI}). In the following
 section we present such classification for a special class of
 superpotentials whose dependence on $k$ is defined by
 terms proportional to $k$ and $\frac1k$ only.

\section{Irreducible matrix superpotentials\label{reducibility}}

To generalize (\ref{s4}) we consider  superpotentials of  the
following special form
\begin{equation}
\label{SP} W_k=kQ +\frac1k R+P,
\end{equation}
where $P$, $R$ and $Q$  are $n\times n$ Hermitian matrices depending
on $x$. Moreover, we suppose that $Q=Q(x)$ is proportional to the
unit matrix. This supposition can be motivated by two reasons:
\begin{itemize}
\item  our goal is to generalize superpotential (\ref{s4}) in which
the term linear in $k$ is proportional to the unit matrix;
\item  restricting ourselves to such $Q$ it is possible to make a
complete classification of the corresponding superpotentials
(\ref{SP}) satisfying shape invariance condition (\ref{SI}).
\end{itemize}

 We do not make any {\it a priori} supposition about
possible values of the continuous independent variable $x$. However
we suppose that relations (\ref{SI}) are valid also for $k\to k'$
where $k'=k+\alpha, k+2\alpha, ... , k+n\alpha$ and $n$ is a natural
number which is  sufficiently large to make the following
speculations.

It is reasonable to restrict ourselves to the case when the matrices
$P$ and $R$ cannot be simultaneously transformed to a block diagonal
form since if such (unitary) transformation is admissible, the
related superpotentials are completely  reducible. Thus we suppose
that the pair of matrices $<P, R>$ is irreducible. Let us show that
in this case it is sufficient to consider $1\times1$ and $2\times2$
matrices only.

Considering the special case $\alpha=0$ we conclude that the
corresponding $W_k$ should be linear in $x$ provided relation
(\ref{SI}) is satisfied:
\begin{gather}W_k=\frac12C_kx+M_k\label{a=0}\end{gather}
where $C_k$ is a constant multiplied by the unit matrix and $M_k$ is
a constant hermitian matrix which can be diagonalized. In this way
we obtain a direct sum of shifted one dimensional oscillators whose
irreducible components can be represented in form (\ref{a=0}) where
$C_k$ and $M_k$ are constants.

Let $\alpha\neq0$. Substituting (\ref{SP}) into (\ref{SI}) and
multiplying the obtained expression by $k^2(k+\alpha)^2$ we obtain
\begin{gather}
\label{fiequation}
AB^2( Q'-\alpha Q^2)+2B^2(P'-\alpha{QP})+\alpha B\{R,P\}+ABR'+\alpha AR^2=B^2C_k
\end{gather}
where $\{R,P\}=RP+PR$ is anticommutator of matrices $P$ and $Q$,
$A=2k+\alpha$, $B=k(k+\alpha)$ and the prime denotes derivative
w.r.t. $x$.

 All terms in the l.h.s. of equations (\ref{fiequation}) are polynomials
 in discrete variable $k$. In order for this equation be consistent,
its r.h.s. (which includes an arbitrary element $C_k$) should also
be a polynomial of the same order whose general form is
\begin{gather}
\label{Ck}
B^2C_k=\nu\alpha AB^2-2\mu B^2-\alpha\lambda B+\rho AB+\alpha\omega^2 A
\end{gather}
where the Greek letters denote
arbitrary parameters.  Substituting (\ref{Ck}) into
(\ref{fiequation}) and equating coefficients for linearly
independent terms we obtain the following system:
\begin{gather}
Q^2\alpha-Q'+\nu\alpha=0,\label{a0}\\\label{a00} P'-\alpha
QP+\mu=0,\\\label{a01} \{R,P\}+\lambda=0\\R'=\rho,\quad
R^2=\omega^2.\label{a8}
\end{gather}

It follows from (\ref{a8}) that $\rho=0$ and $R$ is a constant
matrix whose square is proportional to the unit one.

If $R$ is proportional  to the unit matrix $I$ or is the zero matrix
(in the last case $\omega=0$) the corresponding superpotential
(\ref{SP}) is reducible. Let $\omega\neq0$ and $R\neq \pm\omega I$
then the general form of $P$ satisfying (\ref{a01}) is
\begin{gather}\label{Pgen}P=\frac\lambda{2\omega}R+\tilde P\end{gather} where $\tilde P$ is a matrix
which  anticommutes with $R$.

A straightforward analysis of equation (\ref{a00}) shows that it is
easily integrable, but to obtain non-trivial $Q$ it is necessary to
set $\mu=\lambda=0$. Indeed, without loss of generality hermitian
matrix $R$ whose square is proportional to the unit matrix can be
chosen in the diagonal form:
\begin{gather}\label{R} R=\omega\left(\begin{array}{ll}I_{m\times
m}&0_{m\times s}\\0_{s\times m}&-I_{s\times s}
\end{array}\right),\quad m+s=n\end{gather}
where $\omega\neq 0$ is a constant, $I_{...}$ and $0_{...}$ are the
unit and zero matrices whose dimension is indicated in subindices,
and  without loss of generality we suppose that $s\geq m$.

The corresponding matrix $\tilde P$ satisfying (\ref{8}) has the
following generic form:
\begin{gather}\label{PMM}\tilde P=\left(\bea{ll}0_{m\times
m}&M_{m\times s}\\M^\dag_{s\times m}&\ 0_{s\times
s}\eea\right)\end{gather} where $M_{m\times s}$ is an
arbitrary matrix of dimension ${m\times s}$. Substituting
(\ref{Pgen})--(\ref{PMM}) into (\ref{a00}) we obtain the following
equations:
\begin{gather}\label{rere}\tilde P'=2\alpha Q\tilde P,\\
\left(\frac\lambda{\omega}R+\mu I_{n\times n}\right)Q=0.\label{re3}
\end{gather}

Analyzing equation (\ref{re3}) we conclude that for
$\lambda^2+\mu^2\neq0$ the matrix in  brackets is invertible
and so we have to set $Q=0$. If $Q$ is nontrivial we have to set
$\lambda=\mu=0$. As a result the system (\ref{a00})--(\ref{a8}) is
reduced to the following form
\begin{gather}
Q^2\alpha-Q'+\nu\alpha=0, \label{0} \\
P= \tilde P\exp\left(\alpha\int\!\!Qdx\right), \label{01} \\
\{\tilde P,R\}=0,\quad R^2=\omega^2, \label{00} \\
C_k=\frac{\alpha\omega^2(2k+\alpha)}{k^2(k+\alpha)^2}+\nu\alpha(2k+\alpha) \label{8}
\end{gather}
where both $R$ and $\tilde P$ are constant
matrices.

Thus the problem of classification of matrix valued shape invariant
potentials (\ref{SP}) is reduced to solving  the first order
differential equation (\ref{0}) for function $Q$ and the algebraic
 problem (\ref{00}) for hermitian  matrices $R$ and $\tilde P$.

 Let us show that  hermitian $n\times n$ matrices $\tilde P$ and $R$ which
 satisfy conditions (\ref{00}) can be simultaneously transformed to
 a block diagonal form. Moreover,
 irreducible matrices satisfying (\ref{00}) are nothing but the $2\times2$
 Pauli matrices multiplied by constants, and "$1\times1$ matrices" (scalars)
  satisfying $R\tilde P=0$. Starting with (\ref{R}) and ({\ref{PMM})
  and applying a unitary transformation
  \begin{gather*}R\to R'=URU^\dag,\quad \tilde P\to
  \tilde P'=U\tilde PU^\dag,\\U=\left(\begin{array}{ll}u_{m\times m}&0_{m\times s}
  \\0_{s\times m}&u_{s\times s}\end{array}\right)\end{gather*}
where $u_{m\times m}$ and $u_{s\times s}$ are unitary
submatrices, we obtain
\begin{gather}\label{PMMA}\tilde P'=\left(\bea{ll}0_{m\times
m}&M'_{m\times s}\\M'^\dag_{s\times m}&0_{s\times
s}\eea\right),\quad R'=R\end{gather} with
\begin{gather}\label{PMA}M'_{m\times s}=u_{m\times m}M_{m\times s}
u^\dag_{s\times s}.\end{gather}

Transformation (\ref{PMA}) can be used
to simplify submatrix $M_{m\times s}$. In particular this
submatrix can be reduced to the following form
\begin{gather}\label{diag}M'_{m\times s}=\left({\widetilde M}_{m\times m}\ \
0_{m\times(s-m)}\right)\end{gather} where  $\widetilde M_{m\times
m}$ is a diagonal matrix:
\begin{gather}\label{MM}{\widetilde M}_{m\times
m}=\texttt{diag}(\mu_1,\mu_2,\cdots,\mu_m)\end{gather} where
$\mu_1,\mu_2,...$ are real parameters. Without loss of generality we
suppose that there are $r$ nonzero parameters $\mu_1, \mu_2,...\mu_r
$ with $0\leq r\leq m$ being the rank of matrix $M$.

Notice that transformation (\ref{PMA})--(\ref{MM}) for rectangular
matrices $M$ is called {\it singular value decomposition}.
Such transformations are widely used in linear algebra, see, e.g.,
\cite{diagonalize}.

 But the set of matrices $\{R,\tilde P_A\}$ with $R$
and $\tilde P_A$ given in (\ref{R}) and (\ref{PMMA}), (\ref{diag}),
(\ref{MM}) is completely reducible since by an accordant permutation
of rows and columns they can be transformed to direct sums of
$2\times2$ matrices $\{ R_{2\times2}, {\tilde P}_{2\times2}\}$ where
\begin{gather}\label{reduced}R_{2\times2}=
\left(\bea{cc}\omega& 0\\0&-\omega\eea\right)\equiv\omega\st,\
\  {\tilde P}_{2\times2}=\left(\bea{cc}0&
\mu\\\mu&0\eea\right)\equiv \mu\so,\quad \mu=\mu_1, \mu_2,...,\mu_r\end{gather} and of
$1\times1$ matrices
\begin{gather}\label{reducedS} R_{1\times1}=\pm\omega,\quad {\tilde
P}_{1\times1}=\mu,\quad \mu\omega=0\end{gather} were $\omega$ and $\mu$ are arbitrary
real numbers. The transformation of matrices (\ref{R}) and
(\ref{PMMA}), (\ref{diag}), (\ref{MM})  to the direct sum of
matrices (\ref{reduced}) and (\ref{reducedS}) can be given
explicitly as
\begin{gather*}
R\to URU^\dag,\quad \tilde P\to URU^\dag
\end{gather*}
where $U$ is a unitary matrix whose nonzero
entries  are:
\[
\begin{array}{l}
U_{a\ a}=U_{b\ m+b-1}=U_{m+b\ b+1}=1,  \\
a=1,\ m+s,\ m+ s+1,\ m+ s+2,\ \cdots, n, \quad b=2,\ 3, \cdots,
s+1.
\end{array}
\]

   Thus up to unitary equivalence we have only two versions of
   irreducible matrices $R$ and $P$ which are given
   by equations (\ref{reduced}) and (\ref{reducedS}).

  The remaining equation (\ref{0}) is easily integrable, thus we can
  find all inequivalent irreducible superpotentials (\ref{SP}) in
  explicit form.

\section{Superpotentials and shape invariant potentials\label{SIP}}

There are six different types of solutions of equation (\ref{0}),
namely
\begin{gather}\label{lin1}\begin{split}&Q=0,\ \nu=0, \
\end{split}\end{gather}and\begin{gather}\begin{split}\label{lin2}&
Q=-\frac{1}{\alpha x}, \ \nu=0,\\& Q=-\frac\lambda\alpha,\quad
\nu=-\frac{\lambda^2}{\alpha^2}<0,\\&
Q=-\frac\lambda\alpha\tanh\lambda x,\quad
\nu=-\frac{\lambda^2}{\alpha^2}<0,\\&Q=-\frac\lambda\alpha\coth\lambda
x,\quad \nu=-\frac{\lambda^2}{\alpha^2}<0,\\&
Q=\frac\lambda\alpha\tan\lambda x,\quad
\nu=\frac{\lambda^2}{\alpha^2}>0
\end{split}
\end{gather}
that are defined up to translations $x\rightarrow x+c,\ c$ is an
integration constant, and $\alpha$ is supposed to be nonzero.

 The corresponding matrices $\tilde{P}$ are easily
calculated using equations (\ref{01}) and (\ref{reduced}) or
(\ref{reducedS}).

Let us note that using solutions (\ref{lin1}) or solutions
(\ref{lin2}) for scalar $P$ and $R$ given by relations (\ref{reducedS}),
 we simply recover the known list of shape invariant
potentials which is presented, e.g., in \cite{Khare}, see the table
on pages 291-292 (this list includes  also the harmonic oscillator
(\ref{a=0})). We will not present this list here but note that our
approach gives a simple and straightforward way to find it.

Consider the case when $P$ and $R$ are $2\times2$ matrices
(\ref{reduced}). Now solutions (\ref{lin1}) for $Q$ and solutions
with trivial matrices $P$ are not available since they lead to
reducible superpotentials. However, solutions (\ref{lin2}) are
consistent. Substituting (\ref{01}), (\ref{reduced}), (\ref{lin2})
into (\ref{SP}) we
 obtain the
following list of matrix superpotentials
\begin{gather}
\label{inverse}
W_{\kappa,\mu}=\left(\left(2\mu+1\right)\st-2\kappa-1\right)\frac1{2x}+
\frac{\omega}{2\kappa+1}\so,  \quad \mu>-\frac12,
\\
\label{exp} W_{\kappa,\mu}= \lambda\left(-\kappa+ \mu\exp(-\lambda
x)\so-\frac{\omega}{\kappa}\st\right),\\
\label{tan} W_{\kappa,\mu}=\lambda\left(\kappa\tan\lambda x
+\mu\sec\lambda x\st+\frac{\omega}{\kappa}\so\right), \\
\label{cotanh} W_{\kappa,\mu}=\lambda\left(-\kappa \coth\lambda x+
\mu\csch\lambda x\st-\frac{\omega}{\kappa}\so\right),\quad \mu<0,\
\omega>0,\\
W_{\kappa,\mu}=\lambda\left(-\kappa \tanh\lambda x+ \mu\sech\lambda
x\so-\frac{\omega}{\kappa}\st \right),\label{tanh}
\end{gather}
where we introduce the normalized parameter
$\kappa=\frac{k}{\alpha}.$ These superpotentials are defined up to
translations $x\rightarrow x+c$,  $\kappa\rightarrow \kappa+\gamma$,
and up to unitary transformations $W_{\kappa,\mu}\to
U_aW_{\kappa,\mu}U_a^\dag$ where $U_1=\so,
U_2=\frac1{\sqrt{2}}(1\pm\ri \sd)$ and $U_3=\st$. In particular
these transformations change signs of parameters $\mu$ and $\omega$ in
(\ref{exp})--(\ref{tanh}) and of $\mu+\frac12$ in (\ref{inverse}),
thus without loss of generality we can set
\begin{gather}\label{muo}\omega>0,\quad \mu>0\end{gather} in all
superpotentials (\ref{exp})--(\ref{tanh}). Zero values of these
parameters are excluded if superpotentials
(\ref{inverse})--(\ref{cotanh}) are irreducible.

Conditions (\ref{muo})  can be imposed also for superpotential
(\ref{cotanh}). To unify some following calculations we prefer to
fix the signs of $\mu$ and $\kappa$ in the way indicated in
(\ref{cotanh}).

Notice that the transformations $k\to k'=k+\alpha$ correspond to the
following transformations for $\kappa$:
\begin{gather}\kappa\to\kappa'=\kappa+1\label{kappa}.\end{gather}

If $\mu=0$ and $\omega=1$ then operator (\ref{inverse}) coincides
with the well known superpotential for PS problem (\ref{s4}), but
for $\mu\neq0$ superpotential (\ref{inverse}) is not equivalent to
(\ref{s4}). The other found superpotentials are new also and make it
possible to formulate consistent, exactly solvable problems for
Schr\"odinger equation with matrix potential. The corresponding
potentials $V_\kappa$ can be found starting with
(\ref{inverse})--(\ref{cotanh}) and using definition
(\ref{pfactorization}). Let us rewrite equation (\ref{pfactorization}) as follows:
\begin{gather}\label{V+C}W_{\kappa,\mu}^2-W'_{\kappa,\mu}={ V}_\kappa=\hat V_\kappa+c_\kappa\end{gather}where $c_\kappa$ is a constant and $\hat V_\kappa$ does not include constant terms proportional to the unit matrix.
As a result we obtain
\begin{gather}
\hat V_\kappa=\left(\mu(\mu+1)+\kappa^2-
\kappa(2\mu+1)\st\right)\frac1{x^2}-
\frac{\omega}{x}\so,\label{pot1}\\
\hat V_\kappa=\lambda^2\left(\mu^2\exp(-2\lambda
x)-(2\kappa-1)\mu\exp(-\lambda x)\so+2\omega
\st\right),\label{pot2}\\
\begin{split}&
\hat V_\kappa=\lambda^2\left((\kappa(\kappa-1)+\mu^2)\sec^2\lambda
x+2\omega\tan\lambda x\so\right.
\\&\left.+\mu(2\kappa-1)\sec\lambda x \tan\lambda
x\st \right), \label{pot3}\end{split}\\
\begin{split}& \hat V_\kappa=\lambda^2\left((\kappa(\kappa-1)+\mu^2)
\csch^2(\lambda x)+ 2\omega\coth\lambda
x\so\right.\\&\left.+\mu(1-2\kappa)\coth\lambda x\csch\lambda x\st
\right),\end{split}\label{pot5}\\
\begin{split}&
\hat V_\kappa=\lambda^2\left((\mu^2-\kappa(\kappa-1))\sech^2\lambda
x+2\omega\tanh\lambda x\st\right.\\&\left. -\mu({2\kappa}-1)
\sech\lambda x \tanh\lambda x\so
 \right).
\end{split}
\label{pot4}
\end{gather}
Potentials (\ref{pot1}), (\ref{pot2}), (\ref{pot3}) (\ref{pot5}) and
(\ref{pot4})
 are generated by superpotentials (\ref{inverse}),
(\ref{exp}), (\ref{tan}), (\ref{cotanh})   and (\ref{tanh})
respectively. The corresponding constants $c_\kappa$ in (\ref{V+C})
are
\begin{gather}\label{c1}c_\kappa=\frac{\omega^2}{(2\kappa+1)^2}
\end{gather}
 for potential
(\ref{pot1}),
\begin{gather}\label{c2}c_\kappa=\lambda^2\left(\kappa^2+\frac
{\omega^2}{\kappa^2}\right)\end{gather} for potentials (\ref{pot2}),
(\ref{pot5}), (\ref{pot4}) and
\begin{gather}\label{c3}c_\kappa=\lambda^2\left(\frac{\omega^2}{\kappa^2}-
\kappa^2\right)\end{gather} for potential (\ref{pot3}).

All the above potentials are shape invariant and give rise to
exactly solvable problems for systems of two coupled Schr\"odinger
equations, i.e., for systems of Schr\"odinger-Pauli type.

\section{Dual shape invariance \label{dsi}}

To find potentials (\ref{pot1})--(\ref{pot5}) we ask for their shape
invariance w.r.t. shifts of parameter $\kappa$. The shape invariance
condition together with the supposition concerning the generic form
(\ref{SP}) of the corresponding superpotential make it possible to
define these potentials up to arbitrary parameters $\lambda, \omega,
\kappa$ and $\mu$.

Starting with superpotentials (\ref{inverse})--(\ref{cotanh}) we can
find the related potentials (\ref{pot1})--(\ref{pot5}) in a unique
fashion. But let us consider the inverse problem: to find possible
superpotentials corresponding to given potentials wich in our case
are given by equations (\ref{pot1})--(\ref{pot5}). The problems of
this kind are very interesting since their solutions can be used to
generate families of isospectral hamiltonians. It happens that in
the case of matrix superpotentials everything is much more
interesting since there exist additional superpotentials compatible
with the shape invariance condition.

To find the mentioned additional superpotentials we use the
following observation: potentials (\ref{pot1}), (\ref{pot3}) and
(\ref{pot5}) are invariant with respect to the simultaneous change
\begin{gather}
\label{change}
\mu \to \kappa-\frac12,\quad \kappa\to\mu+\frac12.
\end{gather}
In addition, there exist another transformations of $\mu$ and $\kappa$ but they
lead to the same results.

Thus in addition
to the shape invariance w.r.t. shifts of $\kappa$ potentials
(\ref{pot1}), (\ref{pot3}) and (\ref{pot5}) should be shape
invariant w.r.t. shifts of parameter $\mu$ also. In other words,
superpotentials in Section 5, should be
considered together with  superpotentials which can be obtained from
(\ref{inverse}), (\ref{tan}) and (\ref{cotanh}) using the change
(\ref{change}).

Thus,  we also can represent
potentials (\ref{inverse}),  (\ref{tan}) and (\ref{cotanh}) in the following form
\begin{gather}\label{SS}\widetilde W_{\mu,\kappa}^2-\widetilde W'_{\mu,\kappa}={\hat  V}_{\mu}+c_\mu\end{gather}
where $\hat V_\mu=\hat V_\kappa$, and\begin{gather}\label{SS1}\widetilde
W_{\mu,\kappa}=\frac{\kappa\st-\mu-1}{x}
+\frac{\omega}{2(\mu+1)}\so,\quad
c_\mu=\frac{\omega^2}{4(\mu+1)^2}\end{gather} for $\hat V_k$ given by
equation (\ref{pot1}),
\begin{gather}\label{SS2}
\widetilde W_{\mu,\kappa}=\frac\lambda2\left((2\mu+1)\tan\lambda x
+(2\kappa-1)\sec\lambda x\st+\frac{4\omega}{2\mu+1}\so\right)
\end{gather} for potential (\ref{pot3}), and
\begin{gather}\label{SS3}
 \widetilde W_{\mu,\kappa}=\frac\lambda{2}\left(-(2\mu+1) \coth\lambda
x+ (2\kappa-1)\csch\lambda
x\st-\frac{4\omega}{2\mu+1}\so\right)\end{gather} for potential
(\ref{pot5}).   The related constant
 constant $c_\mu$  is:
\begin{gather}\label{77}
c_\mu=\lambda^2\left(\pm\frac14(2\mu+1)^2+
\frac{4\omega^2}{(2\mu+1)^2}\right)\end{gather}
where the sign "$+$" and "$-$" corresponds to the cases (\ref{SS2}) and (\ref{SS3}) respectively.

We stress that superpartners of potentials (\ref{SS}) constructed
using superpotentials $\widetilde W_{\mu,\kappa}$, i.e.,
\begin{gather}
\label{SSS}
V^+_{\mu} =\widetilde W_{\mu,\kappa}^2+\widetilde W'_{\mu,\kappa}
\end{gather}
satisfy the shape invariance condition since
\begin{gather*}
V^+_{\mu}=V_{\mu+1}+C_\mu
\end{gather*}
with $C_\mu=c_{\mu+1}-c_\mu$.

Thus  potentials (\ref{inverse}), (\ref{tan}) and
(\ref{cotanh}) admit a dual supersymmetry, i.e., they are shape
invariant w.r.t. shifts of two parameters, namely, $\kappa$ and
$\mu$. More exactly, superpartners for  potentials (\ref{pot1}),
(\ref{pot3}) and (\ref{pot5}) can be obtained  either by shifts of
$\kappa$ or by shifts of $\mu$ while simultaneous shifts are
forbidden. We call this phenomena {\it dual shape invariance}.

Notice that the remaining potentials (\ref{pot2}) and (\ref{pot4})
do not posses the dual shape invariance in the sense formulated
above. In potential (\ref{pot2})  parameter $\mu$ is not essential.
It is supposed to be non-vanishing (since for $\mu=0$ the
corresponding superpotential is reducible) and can be normalized to
the unity by shifting  independent variable $x$.

The hamiltonian with potential (\ref{pot4}) is not invariant w.r.t.
change (\ref{change}). However if we suppose that parameter $\mu$ be
purely imaginary, i.e., set $\mu= \ri \tilde \mu$ with $\tilde \mu$
real, the corresponding potential admits discrete symmetry
(\ref{change}) for parameters $\kappa$ and $\tilde\mu$ and thus
possesses dual supersymmetry with shape invariance. In this way we
obtain a consistent model of "PT-symmetric quantum mechanics
\cite{bend}"
 with the dual shape invariance. Discussion of this
model lies out  the scope  of  present paper. We only note that for
$\omega=0$ the corresponding potential is decoupled to a direct sum
of potentials discussed in \cite{znojl}.

\section{Exactly solvable problems\\ for systems
 of
Schr\"odinger equations\label{SE}}

Consider the Schr\"odinger equations
\begin{gather}
\hat H_\kappa\psi\equiv\left(-\frac{\p^2}{\p x^2}+\hat V_\kappa\right)
\psi=E_\kappa\psi\label{se2}
\end{gather}
where $\hat H_\kappa=a^+_{\kappa,\mu}a^-_{\kappa,\mu}+c_\kappa$ and $\hat V_\kappa$ are matrix
potentials represented in (\ref{pot2})--(\ref{c2}).
 Since all these potentials
 are shape invariant, equations (\ref{se2}) can be integrated using
 the standard technique of SSQM. An algorithm for construction of
 exact solutions of supersymmetric a shape invariant Schr\"odinger equations
 includes the following steps (see, e.g., \cite{Khare}):
 \begin{itemize}
 \item To find the ground state solutions $\psi_0(\kappa,\mu,x)$ which are proportional to
  square integrable solutions of the first order equation
  \begin{gather}
  \label{psi0}
  a_{\kappa,\mu}^-\psi_0(\kappa,\mu,x)\equiv
  \left( \frac{\p}{\p
  x}+W_{\kappa,\mu}\right)\psi_0(\kappa,\mu,x)=0.
  \end{gather}
  In view of (\ref{V+C}) function $\psi_0(\kappa,\mu,x)$ solves equation
(\ref{se2}) with
\begin{gather}
E_\kappa=E_{\kappa,0}=-c_\kappa\label{E0k}.
\end{gather}
\item To find a solution $\psi_1(\kappa,\mu,x)$ for the first excited state which is
defined by the following relation:
\begin{gather}
\label{psi1}
\psi_1(\kappa,\mu,x)=a^+_{\kappa,\mu}
\psi_0(\kappa+1,\mu,x)\equiv\left(- \frac{\p}{\p
x}+W_{\kappa,\mu}\right)\psi_0(\kappa+1,\mu,x).
\end{gather}
 Since $a_\kappa^\pm$ and $\hat{H}_\kappa$ satisfy the interwining relations
  \[\hat H_\kappa a_{\kappa,\mu}^+=a_{\kappa,\mu}^+ \hat H_{\kappa+1}\]
  function (\ref{psi1})  solves
  equation (\ref{se2}) with $E_\kappa=E_{\kappa,1}=-c_{\kappa+1}$.
  \item Solutions for the second excited state can be found as
  $\psi_2(\kappa,\mu,x)=a^+_{\kappa,\mu}\psi_1(\kappa+1,\mu,x)$,
  etc. Finally, solutions
  which correspond to $n^{th}$ exited state for any admissible natural number $n>0$
  can be represented as
\begin{gather}\label{psin}\psi_n(\kappa,\mu,x)=
a_{\kappa,\mu}^+a_{\kappa+1,\mu}^+ \cdots
a_{\kappa+n-1,\mu}^+\psi_0(\kappa+n,\mu,x).
\end{gather} The corresponding eigenvalue $E_{\kappa,n}$ is equal to
$-c_{\kappa+n}$.
\item For systems admitting the dual shape invariance it is
necessary to repeat the steps enumerated above using alternative (or additional)
superpotentials.
\end{itemize}

All potentials presented in the previous section generate integrable
models with Hamiltonian (\ref{se2}). However, it is desirable  to
analyze their consistency. In particular, it is necessary to verify
that there exist square integrable solutions of equation
(\ref{psi0}) for the ground state.

In the following sections we prove that such solutions exist for all
superpotentials given by equations (\ref{inverse})--(\ref{cotanh})
and (\ref{SS1})--(\ref{SS3}). We will see that to obtain
normalizable ground state solutions it is necessary to impose
certain conditions on parameters of these superpotentials.

To finish this section we present energy spectra for models
(\ref{se2}) with potentials (\ref{pot1})--(\ref{pot5}):
\begin{gather}\label{EV0}E=
-\frac{\omega^2}{(2N+1)^2}\end{gather}
for potential (\ref{pot1}),
\begin{gather}\label{EV1}E=-\lambda^2\left(N^2+\frac{\omega^2}
{N^2}\right)\end{gather} for potentials (\ref{pot2}), (\ref{pot4}),
(\ref{pot5}), and
\begin{gather}\label{EV2}E=\lambda^2\left(N^2-\frac{\omega^2}{N^2}\right)\end{gather}
for potentials  (\ref{pot3}).

In equations (\ref{EV0})--(\ref{EV0}) we omit subindices labeling
the energy levels. The spectral parameter $N$ can take the following
values
\begin{gather}\label{Nn}N=n+\kappa,  \end{gather} and (or)
\begin{gather}\label{Nnm}N=n+\mu+\frac12  \end{gather}
where $n=0,1,2,...$ are natural numbers which can take any values
for potentials (\ref{pot1})--(\ref{pot3}). For potentials
(\ref{pot2}), (\ref{pot4}) and (\ref{pot5}) {\it with a fixed $k<0$}
the admissible values of $n$ are bound by the condition
$(k+n)^2>|\omega|$, see section 9.

For potential (\ref{pot2}) the spectral parameter is defined by
equation (\ref{Nn}). For  potentials (\ref{pot1}), (\ref{pot3}),
(\ref{pot4}) the form of $N$ depends on relations between parameters
$\kappa$ and $\mu$, see section 9.

\section{Some special values of parameters and isospectrality\label{iso}}

Let us show that for some values of parameters $\mu$ and $\kappa$
potentials (\ref{pot1})--(\ref{pot4}) are isospectral with direct
sums of known scalar potentials.

Considering potential (\ref{pot1}) and using its dual shape
invariance it is possible to discover  that for  half integer $\mu$
$V_\kappa$ can be transformed to a direct sum of scalar Coulomb
potentials. Indeed, its superpartner obtained with using
superpotential (\ref{SS}) with opposite sign, i.e., $\hat
W_{\mu,\kappa}=-\widetilde W_{\mu,\kappa}$   looks as:
\begin{gather}
\label{CC}
\hat{V}_{\kappa,\mu}^+=\hat W^{2}_{\mu,\kappa}+{\hat
W}'_{\mu,\kappa}+c_\mu= \left(\mu(\mu-1)+\kappa^2-
\kappa(2\mu-1)\st\right)\frac1{x^2}-
\frac{\omega}{x}\so.
\end{gather}

Considering $\hat V_{\kappa,\mu}^+=\hat V_{\kappa,\mu+1}$ as the
main potential and calculating its superpartner with using
superpotential $\hat W_{\mu+1,\kappa}$ we come to equation
(\ref{CC}) with $\mu\to\mu-2$, etc. It is easy to see that
continuing this procedure we obtain on some step the following
result:
\begin{gather}  \hat
V_{\kappa,\tilde\mu}^+= \frac{l(l+1)}{x^2}-
\frac{\omega}{x}\so,\quad l=\kappa-\frac12\label{pot11}\end{gather}
where $\tilde\mu=\mu+n,\ n=-\mu-\frac12$. Diagonalizing matrix
 $\so\to\st$ we reduce (\ref{pot11}) to a
direct sum of attractive and repulsive Coulomb potentials written in
radial variables. It means that for  negative and half integer $\mu$
our potential (\ref{pot1}) is isospectral with the Coulomb one.

In analogous way we can show that potentials (\ref{pot3}) with
 half integer $\kappa$ or  integer $\mu$ is
isospectral with the potential
\begin{gather}\hat V_\kappa=\lambda^2\left(r(r-1)
\sec^2\lambda x+2\omega\tan\lambda x\so\right), \quad
r=\frac12\pm\mu\quad  \texttt{or}\quad r=\kappa,
\label{pot33}\end{gather} which is equivalent to the direct sum of
two trigonometric Rosen-Morse potentials. Under the same conditions
for parameters $\mu$ and $\kappa$ potential (\ref{pot4}) is
isospectral with the following potential:
\begin{gather}\hat V_\kappa=\lambda^2\left(r(r-1)
\csch^2(\lambda x)+ 2\omega\coth\lambda x\so \right)\label{pot55}
\end{gather} which is equivalent to the direct sum of two Eckart
potentials. Finally, potential (\ref{pot4}) is isospectral with
\begin{gather}\hat
V_\kappa=\lambda^2\left(r(r-1)\sech^2\lambda x+2\omega\tanh\lambda
x\st
 \right),\quad r=\frac12\pm\sqrt{\mu^2+\frac12}\quad
\label{pot44}\end{gather} provided $\kappa$ is negative half
integer. Potential (\ref{pot44}) is equivalent to the direct sum of
two hyperbolic Rosen-Morse potentials.

Thus for some special values of parameters $\mu$ and $\kappa$ we can
establish the isospectrality relations of matrix potentials
(\ref{pot1})--(\ref{pot4}) with well known scalar potentials. This
observation is supported by the direct comparison of spectra
(\ref{EV0})--(\ref{EV0}) with the spectra of Schr\"odinger equation
with Coulomb, Rosen-Morse and Eckart potentials which can be found,
e.g., in \cite{Khare}.

Let us note that setting in (\ref{pot2})--(\ref{pot4}) $\omega=0$ we
also come to the direct sums of shape invariant potentials, namely,
Morse, Scraft and generalized P\"oshl-Teller ones. However for
nonzero $\omega=0$ and $\mu$, $\kappa$ which do not satisfy
conditions imposed to obtain (\ref{pot11})--(\ref{pot44}) the found
potentials cannot be transformed to the mentioned directs sums using
the consequent Darboux transformations.

\section{Ground state solutions\label{GSS}}

 Let us find the ground state solutions for equations (\ref{se2}) with
 shape invariant potentials (\ref{pot1})--(\ref{pot5}). To do this it is necessary
 to solve equations (\ref{psi0}) where $W_{\kappa,\mu}$ are
 superpotentials given in (\ref{inverse})--(\ref{cotanh}),
 and analogous equation with superpotentials (\ref{SS1})--(\ref{SS3}).
 The corresponding solutions are square integrable two component functions which we denote as:
\begin{gather}\psi_0(\kappa,\mu,x)=\left(\bea{c}\varphi\\
\xi\eea\right)\label{psi00}.\end{gather}

In this section we find the ground state solutions considering
 consequently all the mentioned potentials.

\subsection{Ground states for systems with potentials (\ref{pot1})
and (\ref{pot2})}

 Let us start
with the superpotential defined by equation (\ref{inverse}).
Substituting (\ref{inverse}) and (\ref{psi00}) into (\ref{psi0}) we
obtain the following system:
\begin{gather}\label{GS11}
\frac{\p \varphi}{\p
x}+\left(\mu-\kappa\right)\frac{\varphi}{x}+\frac\omega{2\kappa+1}\xi=0,
\\\label{GS12}
\frac{\p \xi}{\p
x}-\left(\mu+\kappa+1\right)\frac{\xi}{x}+\frac\omega{2\kappa+1}\varphi=0.\end{gather}
Solving (\ref{GS12}) for $\varphi$, substituting the solution into
(\ref{GS11}) and making the change
\begin{gather}\label{bes}\xi=y^{\kappa+1}\hat\xi(y),\ \ \ y=\frac{\omega
x}{2k+1}\end{gather} we obtain the equation
\begin{gather}\label{bessel}y^2\frac{\p^2 \hat\xi}{\p y^2}+
y\frac{\p \hat\xi}{\p y}-
\left(y^2+\mu^2\right)\hat\xi=0.\end{gather} Its solution is a
linear combination of modified Bessel functions:
\begin{gather}\label{GS13}\hat\xi=C_1K_{\mu}(y)+C_2 I_{\mu}(y).
\end{gather}

To obtain a square integrable solution we have to set in
(\ref{GS13}) $C_2=0$ since $I_{\mu}(y)$ turns to infinity with $x\to
\infty$. Then substituting  (\ref{GS13}) into (\ref{bes}) and using
(\ref{GS12}) we obtain solutions for system (\ref{GS11}),
(\ref{GS12}) in the following form:
\begin{gather}\label{GS1}\varphi=y^{\kappa+1}
K_{\mu+1}(y), \quad \xi=y^{\kappa+1} K_{|\mu|}(y)
\end{gather} where $y$ is the variable defined in
(\ref{bes}), $\omega x/(2\kappa+1)\geq0$.

 Functions (\ref{GS1}) are square integrable provided
parameter $\kappa$ is positive and satisfies the following relation:
\begin{gather}\quad \kappa-\mu>0\label{condk1}.\end{gather}

If this condition is violated, i.e.,
\begin{gather}\label{SS666}\kappa-\mu\leq0\end{gather}
we cannot find ground state vector using equation (\ref{psi0}) with
superpotential (\ref{inverse}) since its solutions  (\ref{GS1}) are
not square integrable. But since potential (\ref{pot1}) admits the
dual shape invariance, it is possible to make an alternative
factorization of equation (\ref{se2}) using superpotential
(\ref{SS1}) and search for normalizable solutions of the following
equation:
\begin{gather}\tilde a_{\mu,\kappa}^-\tilde\psi_0(\mu,\kappa,x)\tilde\psi_0(\mu,\kappa,x)=0.\label{psi0m}\end{gather}
   where (and in the following)
\begin{gather}\label{amu}\tilde a^-_{\mu,\kappa}= \frac{\p}{\p
  x}+\widetilde W_{\mu,\kappa},\quad \tilde a^+_{\mu,\kappa}= -\frac{\p}{\p
  x}+\widetilde W_{\mu,\kappa}.\end{gather}
   Indeed, solving (\ref{psi0m})
   we obtain a perfect ground state vector:
  \begin{gather}\label{SS10}\tilde\psi_0(\mu,\kappa,x)=
  \left(\bea{c}\tilde\varphi\\
\tilde\xi\eea\right),\quad \tilde\varphi=y^{\mu+\frac32}
K_{|\nu|}\left(y\right), \quad\tilde\xi=
y^{\mu+\frac32}K_{|\nu-1|}\left(y\right)
\end{gather} where $y=\frac{\omega x}{2(\mu+1)}$ and  $\nu=\kappa+1/2.$
The normalizability conditions for solution (\ref{SS10}) are:
\begin{gather}\kappa-\mu<1,\quad \texttt{if}\quad \kappa\geq0\label{SS5}\end{gather}
and
\begin{gather}\kappa+\mu>1,\quad \texttt{if}\quad
\kappa<0.\label{SS55}\end{gather}

 It is important to note that conditions (\ref{condk1}) and
(\ref{SS5}) are compatible provided
\begin{gather}\label{SS6}\kappa>0,\quad 0<\kappa-\mu<1.\end{gather}
 Conditions (\ref{condk1}) and (\ref{SS55}) are incompatible.

Thus if parameters $\mu$ and $\kappa$ satisfy (\ref{SS666}) and
(\ref{SS5}) , equation (\ref{se2}) admits
 ground state solutions (\ref{SS10}). If (\ref{condk1}) is satisfied but
(\ref{SS5})  is not true, the ground state solutions are given
 by relations (\ref{GS1}). If
 condition (\ref{SS6}) is satisfied both solutions  (\ref{GS1}) and (\ref{SS10})
 are available. In the special case $\kappa=\mu+1/2$ solutions (\ref{GS1}) and
 (\ref{SS10}) coincide.

 Notice that our convention that parameter $\mu$ is positive
 excludes  the case $\mu=-1/2$ when potential (\ref{pot1}) is reduced
 to a direct sum of Coulomb potentials, see section \ref{iso}.

Analogously, considering equation (\ref{psi0}) with superpotential
(\ref{exp}) and representing its solution in the form (\ref{psi00})
with
\begin{gather*}\xi=y^{\frac12-\kappa}\hat\xi(y),\quad
\varphi=y^{\frac12-\kappa}\hat\varphi(y), \quad  y=\mu\exp(-\lambda
x)\end{gather*}
 we find the following solutions: \begin{gather}\label{GS4}
\varphi=y^{\frac12-\kappa}K_{|\nu|}( y),\quad
\xi=-y^{\frac12-\kappa}K_{|\nu-1|}(y)\end{gather}where
$\nu=\omega/\kappa+1/2$ and parameters $\omega$ and $\kappa$ should
satisfy the conditions
\begin{gather}  \kappa<0,\quad \kappa^2>\omega.\label{th1}
\end{gather}

 Since potential (\ref{pot2}) does not admit the dual shape
 invariance, there are no other ground state solutions.

\subsection{Ground states for systems with potentials
(\ref{pot3})--(\ref{pot4})}

 In analogous manner we find solutions of equations (\ref{psi0}) and
  (\ref{psi0m}) for the remaining superpotentials
  (\ref{exp})--(\ref{cotanh}). Let us present them without
  calculational details.

  Solving equation (\ref{psi0}) for superpotential (\ref{tan}) we
obtain two normalizable solutions, the first of which is:
\begin{gather}\label{GS3}\begin{split}&\varphi_1=
y^{\frac{\kappa-\mu}2}(1-y)^{\frac{\kappa+\mu}2} {_2F_1}\left(a,b,c;
y\right),\\&\xi_1=\frac{2\omega}{\kappa(2\mu-1)}
y^{\frac{1+\kappa-\mu}2}(1-y)^{\frac{1+\kappa+\mu}2}
{_2F_1}\left(a+1,b+1,c+1; y\right).\end{split}\end{gather}  Here
$_2F_1(a,b,c;y)$ is the hypergeometric function,
\begin{gather}\begin{split}&\label{tan11}a=-\ri\frac\omega\kappa,\quad
b=\ri\frac\omega\kappa, \quad c=\frac12-\mu\neq0, \\& y=\frac12(\sin
\lambda x+1),\quad -\frac\pi2\leq\lambda
x\leq\frac\pi2,\end{split}\end{gather} and parameters $\mu$ and
$\kappa$ are constrained by the conditions (\ref{condk1}) and
\begin{gather}\label{condk2}\kappa+\mu>0.\end{gather}

The second solution is
\begin{gather}\label{tan2}\begin{split}&\varphi_2=
y^{\frac{1+\kappa+\mu}2}(1-y)^{\frac{\kappa+\mu}2}
{_2F_1}\left(a,b,c;
y\right),\\&\xi_2=-\frac{(2\mu+1)\kappa}{2\omega}\left(\frac{1-y}y\right)^\frac12\varphi_2
\\&-\frac{\kappa^2(2\mu+1)^2+4\omega^2}{2\omega\kappa(2\mu+3)}
y^\frac{2+\kappa+\mu}{2}(1-y)^{\frac{1+\kappa+\mu}2}
{_2F_1}\left(a+1,b+1,c+1; y\right)\end{split}\end{gather} where
variable $y$ is the same as in (\ref{GS3}),
\begin{gather}\label{tan4}a=\mu+\frac12-\ri\frac\omega\kappa,\quad
b=\mu+\frac12+\ri\frac\omega\kappa,\quad c=\mu+\frac32,\end{gather}
and parameters $\kappa, \mu$ again should satisfy conditions
(\ref{condk1}) and (\ref{condk2}).

Using the dual shape invariance of potential (\ref{pot3}) we can
find additional (or alternative) ground state solutions using
equation (\ref{psi0m}) with superpotential (\ref{SS2}). In this way
we obtain
\begin{gather}\label{GS3d}\begin{split}&\tilde\varphi_1=
y^{\frac{\mu-\kappa+1}2}(1-y)^{\frac{\kappa+\mu}2}
{_2F_1}\left(a,b,c;
y\right),\\&\tilde\xi_1=\frac{4\omega}{(2\kappa-3)(2\mu+1)}
y^{\frac{2+\mu-\kappa}2}(1-y)^{\frac{1+\kappa+\mu}2}
{_2F_1}\left(a+1,b+1,c+1; y\right).\end{split}\end{gather}  Here
 variable $y$ and its domain are the same as given in (\ref{tan11}),
\begin{gather}\label{tan12}-a=b=\frac{2\ri\omega}{2\mu+1},\quad
 c=1-\kappa, \end{gather}
and parameters $\mu$ and $\kappa$ are constrained by the conditions
(\ref{condk2}) and
\begin{gather}\label{condk3}\kappa-\mu<1.\end{gather}
The second solution is
\begin{gather}\label{tan2d}\tilde\varphi_2=\varphi_2,\quad
\tilde\xi_2=\xi_2\end{gather} where $\varphi_2$ and $\xi_2$ are
functions defined by equation (\ref{tan2}) where arguments $a,b$ and
$c$ of the hypergeometric function differs from (\ref{tan4}) and
have the following form:
\begin{gather*}a=\kappa-\frac{2\ri\omega}{2\mu+1},\quad
b=\kappa-\frac{2\ri\omega}{2\mu+1}\quad
c=\kappa+\frac12,\end{gather*} and parameters $\kappa, \mu$ should
satisfy conditions (\ref{condk2}) and (\ref{condk3}).

Thus for potential (\ref{pot3}) we have three versions of
constraints for parameters $\mu$ and $\kappa:$
\begin{gather}\label{condk4}\kappa-\mu\geq1,\\\kappa-\mu\leq0\label{condk5}\end{gather}
or
\begin{gather}0<\kappa-\mu<1.\label{condk6}\end{gather} In addition,
condition (\ref{condk2}) should be imposed.

For the cases (\ref{condk4}) and (\ref{condk5}) the ground state
solutions are given by equations (\ref{GS3}), (\ref{tan2}) and
(\ref{GS3d}), (\ref{tan2d}) correspondingly while in the case
(\ref{condk6}) all solutions (\ref{GS3}), (\ref{tan2}), (\ref{GS3d})
and (\ref{tan2d}) are available.

 For superpotential (\ref{cotanh})
we obtain the following solution of equation (\ref{psi0}):
\begin{gather}\label{GS21}\begin{split}&
\varphi_1=(1-y^2)^{\frac\omega\kappa-\kappa}y^{\kappa-\mu}{_2F_1}
\left(a,b,c, y^2\right),\\& \xi_1= -y\varphi_1
+\frac{a+c}{c}y^{\kappa-\mu+1}
(1-y^2)^{1-\kappa+\frac\omega\kappa}{_2F_1}\left(a+1,b+1,c+1;
y^2\right)
\end{split}\end{gather}
where
 \begin{gather}a=\frac\omega\kappa,\quad b=a+c,\quad c=\frac12-\mu, \quad
 y=\tanh \frac{\lambda x}2\label{cth}\end{gather}
and parameters $\kappa, \mu, \omega$ satisfy conditions
(\ref{condk1}) and (\ref{th1}).
 These conditions are compatible iff $\mu<0$.

One more solution for superpotential (\ref{cotanh}) has components
given below:
\begin{gather}\label{cotanh2}\begin{split}&\xi_2=
(1-y^2)^{\frac\omega\kappa-\kappa}y^{\kappa-\mu+1}{_2F_1}\left(a,b,c;
y^2\right),\\& \varphi_2=\left(\frac{\kappa(2\mu-1)(y^2-1)}{2\omega
y}-y\right)\varphi_2\\&+ \frac{\kappa ab}{\omega c}
(1-y^2)^{1-\kappa+\frac\omega\kappa}y^{\kappa-\mu+2}{_2F}_1(a+1,b+1,c+1;y^2)\end{split}\end{gather}
where $y$ is the variable given in (\ref{cth}),
\begin{gather}\label{cotanh22}a=1+\frac\omega\kappa,
\quad b=c-a,\quad c=\frac32-\mu.\end{gather} Solution
(\ref{cotanh2}) is normalizable provided conditions (\ref{condk1})
and (\ref{th1}) are satisfied.

Potential (\ref{pot5}) possesses the dual shape invariance thus we
also should  find solutions of equation (\ref{psi0m}) with
superpotential (\ref{SS3}). The explicit expression of these
solutions can be obtained from (\ref{GS21}) and (\ref{cotanh2})
using the change (\ref{change}).  To obtain consistent solutions the
additional conditions
\begin{gather}\label{condmu}\mu<0, \quad \left(2\mu+1\right)^2>4\omega>0\end{gather}
should be imposed instead of  (\ref{th1}).

   For superpotential (\ref{tanh}) we also have two
ground state vectors which solve equation (\ref{psi0}). The first of
them has the following components:
\begin{gather}\label{GS2}\begin{split}&
\varphi_1=y^{-\frac{\kappa}2+ \frac\omega{2\kappa}}
(1-y)^{-\frac{\kappa}2- \frac\omega{2\kappa}}{_2F_1}\left(a,b,c;
y\right),\\& \xi_1=-\frac{2\mu \kappa}{2\omega+\kappa}
y^{\frac12-\frac\kappa2+\frac{\omega}{2\kappa}}(1-y)^{\frac12-\frac\kappa2-
\frac{\omega}{2\kappa}} {_2F_1}\left(a+1,b+1,c+1; y\right),
\end{split}\end{gather}  where the parameters and
variables are
 \begin{gather}a=-\ri\mu,\quad b=\ri\mu,\quad
 c=\frac12+\frac\omega\kappa\neq0,\quad
y=\frac12(\tanh \lambda x+1)\label{th}\end{gather} with $\kappa$ and
$\mu$ satisfying (\ref{th1}).
 The second solution looks as:
\begin{gather}\begin{split}&\varphi_2=y^{\frac12-\frac\kappa2-
\frac\omega{2\kappa}}{(1-y)^{-\frac\kappa2-\frac\omega{2\kappa}}}{_2F}_1(a,b,c;y),\\&
\xi_2=\frac{2\omega-\kappa}{2\kappa\mu}\left(\frac{1-y}y\right)^\frac12\varphi_2\\&-
\frac{(\kappa-2\omega)^2+4\mu^2\kappa^2}{2\mu
\kappa(3\kappa-2\omega)}y^{1-\frac\kappa2-\frac\omega{2\kappa}}
(1-y)^{\frac12-\frac\kappa2-\frac\omega{2\kappa}}
{_2F}_1(a+1,b+1,c+1;y)\end{split}\label{th2}\end{gather}where $y$ is
the variable defined in (\ref{th}),
\begin{gather}\label{th22}a=\frac12-\frac\omega\kappa-\ri \mu,\quad
b=\frac12-\frac\omega\kappa+\ri \mu,\quad
c=\frac32-\frac\omega\kappa,\end{gather} and parameters $\kappa,
\omega$ should satisfy condition (\ref{th1}).

Due to the absence of dual shape invariance of potential
(\ref{pot4}) there are no alternative ground state solutions.

 Thus we find ground state solutions for equation (\ref{se2})
 with all potentials (\ref{pot1})--(\ref{pot5}). These solutions  are square
integrable and
 correspond to the eigenvalues  $E_\kappa=-c_\kappa$ or $E_\mu=-c_\mu$
 where $c_\kappa$ and $c_\mu$ are given by equations (\ref{c1})--(\ref{c3})
 and (\ref{77}).

\section{Exited states}

We already know ground state solutions for Schr\"odinger equations
with potentials given by relations (\ref{pot1})--(\ref{pot5}).
Solutions which correspond to $n^{th}$ energy level can be obtained
starting, e.g.,  with the ground state solutions (\ref{GS1}),
(\ref{GS4}), (\ref{GS3}), (\ref{tan2}), (\ref{GS21}),
(\ref{cotanh2}), (\ref{GS2}), (\ref{th2}) and applying equation
(\ref{psin}). However it is necessary to make sure  that such
defined  ground and exited states
   are square integrable.

Let us first consider  potential (\ref{pot1}) which admits the dual
shape invariance. This invariance enables to make factorization of
the corresponding
 Schr\"odinger equation and find ground state solutions  using either
 superpotential  (\ref{inverse})
or (\ref{SS1}), or even both of them, depending on given initial
values of parameters $\kappa$ and $\mu$. Namely if (\ref{condk1}) is
satisfied but
 (\ref{SS5})  is not true , the ground state solutions are given
 by equations (\ref{psi00}) and (\ref{GS1}) and exited states are
 given by equation (\ref{psin}).

  The corresponding energy levels are given by equations
 (\ref{EV0}) and (\ref{Nn}).

 If parameters $\mu$
and $\kappa$ satisfy (\ref{SS666}) and one of conditions (\ref{SS5})
or (\ref{SS5}), equation (\ref{se2}) admits
 ground state solutions (\ref{SS10}) and exited states are  defined by the following
 equation:
\begin{gather}\label{psinka}\tilde\psi_n(\kappa,\mu,x)=
\tilde a_{\kappa,\mu}^+\tilde a_{\kappa+1,\mu}^+ \cdots \tilde
a_{\kappa+n-1,\mu}^+\tilde\psi_0(\kappa+n,\mu,x)
\end{gather}
where $\tilde a_{\kappa,\mu}^+=-\frac{\p}{\p x}+\widetilde
W_{\mu,\kappa}$ and  $\widetilde W_{\mu,\kappa}$ is the alternative
superpotential given by (\ref{SS1}). The corresponding energy levels
are given by formulae (\ref{EV0}) and (\ref{Nnm})

  If
 condition (\ref{SS6}) is satisfied both versions of
 solutions and energy levels given above
 are available. In the special case $\kappa=\mu+1/2$ solutions (\ref{GS1})
 and
 (\ref{SS10}) coincide.

Let us start with the ground state solution (\ref{GS1}). Its
normalizability  is almost evident since the modified Bessel
function has the only singular point, namely, $y=0$, and decreases
exponentially at infinity. Moreover, at $y=0$ there is the inverse
power singularity, i.e., $K_\nu\sim\frac1{y^\nu}$ with $\nu=\mu $
or $\nu=\mu+1$
which is perfectly
compensated by the multiplier  $y^{\kappa+1}$ provided $\kappa$
satisfies (\ref{condk1}).

Calculating the first exited state (\ref{psi1}) we can use  relation
(\ref{psi0}) where $\kappa\to\kappa+1$, thus
\begin{gather}\label{exit1}\psi_1(k,\mu,y)=
(W_{\kappa+1,\mu}+W_{\kappa,\mu})
\psi_0(\kappa+1,\mu,y).\end{gather} Again we recognize a good
behavior at the singularity point $y=0$ since
$(W_{\kappa+1,\mu}+W_{\kappa,\mu})\sim\frac1y+\cdots$ and
$\psi_0(\kappa+1,\mu,y)\sim y\psi_0(\kappa,\mu,y)$.

Let us suppose that the wave function corresponding to $n^{th}$
exited state is regular at $y=0$ and has the following form
\begin{gather}\label{psinth}\psi_n(\kappa,\mu,y)=
W\psi_0(\kappa+n,\mu,y)\end{gather}
where $\psi_0(\kappa+n,\mu,y)$ is the ground state solution given by
relations (\ref{psi00}), (\ref{GS1}) and $W$ is a matrix depending
on $y$ and $k$.
 Then \begin{gather}\begin{split}&\label{psinn}\psi_{n+1}(\kappa,\mu,y)=
 \left(-\frac{\p}{\p y}+W_\kappa\right)\psi_n(\kappa+n+1,\mu,y)\\&=
 \left(-\frac{\p W}{\p y}+W_\kappa W+WW_{\kappa+n+1}\right)
 \psi_0(\kappa+n+1,\mu,y)\end{split}\end{gather}
 where we use the fact that $\psi_0(\kappa+n+1,\mu,y)$ solves the equation
 (\ref{psi0})
where $\kappa\to\kappa+n+1$.

Using (\ref{psinn}) it is not difficult to show that if function
$\psi_{n}(\kappa+1,\mu,y)$ be square integrable then
$\psi_{n+1}(\kappa,\mu,y)$ is square integrable too. Indeed, at the
neighborhood of the   singularity point $y=0$  the ground state
functions are related as $\psi_0(\kappa+n+1,\mu,y)\sim
y\psi_0(\kappa+n,\mu,y)$, and $\left(-\frac{\p W}{\p y}+W_\kappa
W+WW_{\kappa+1}\right)\sim \frac1yW$. Thus $\psi_{n+1}(\kappa,\mu,y)$ is
regular at $y=0$ provided $\psi_{n}(\kappa,\mu,y)$ be regular. Since for
$n=0,1$ our supposition is fulfilled we conclude by induction that
wave functions $\psi_n(\kappa,\mu,y)$ (\ref{psin}) are normalizable for
any $n$.

By direct repeating the above  speculations we can prove the square
integrability of  ground state vector (\ref{SS10}) and exited state
vectors given by relation (\ref{psinka}). In fact the only thing we
need is to change $\psi_n(\kappa,\mu,y)$ and $W_{\kappa,\mu}$ by their
counterparts $\tilde\psi_n(\kappa,\mu,y)$ and $\tilde W_{\mu,\kappa}$.

 In complete analogy with the above one can
prove the normalizability of the exited states for the case when the
superpotential is given by equation (\ref{exp}). However there is an
essentially new point which is generated by condition (\ref{th1}).
The think is that solutions (\ref{GS4}) being well defined for
$\kappa$ and $\mu$ satisfying conditions (\ref{condk1}) and
(\ref{th1}),   can loose their square integrability  after the
change $\kappa\to\kappa+n$ for a sufficiently large $n$. Namely, in
order to obtain a normalizable solution $\psi_0(\kappa+n,\mu,x)$  we
have to ask for $(\kappa+n)^2\geq\omega>0$. Since $\kappa$ is
negative we have the following restriction for $n$:
\begin{gather}n<|\kappa|-\sqrt{\omega}\label{nko}.\end{gather}
It is possible to show that if $\psi_0(\kappa+n,\mu,x)$ is not
normalizable the same is true for exited states (\ref{psin}).

Let us consider the ground state solutions (\ref{GS3}). This
solution like all the remaining solutions (\ref{tan2}),
(\ref{GS3d}}), (\ref{tan2d}),  (\ref{GS21}), (\ref{cotanh2}),
(\ref{GS2}), (\ref{th2}) is expressed via linear combinations of the
following elements:
\begin{gather}\label{el}y^A(1-y)^B{_2F_1}(a,b,c;y)\end{gather}
where parameters $a, b$ and $c$ are given by equation (\ref{tan11}),
$A=\frac{\kappa-\mu}2, B=\frac{\kappa+\mu}2$ for component
$\varphi_1$, etc.

In accordance with its definition, variable $y$ belongs to the
interval $[0,1]$ and there are two points which are "suspicious
w.r.t. singularity", namely, $y=0$ and $y=1$. In order the solution
to be regular (and equal to zero) in these points it is necessary
and sufficient to ask for $A>0, B>0$ and $\Re e(B+c-a-b)>0$. Exactly
these conditions generate restrictions (\ref{th1}) for parameters
$\kappa$
 which guarantee the solution normalizability. The same is
true for solutions (\ref{tan2}), (\ref{GS2}), (\ref{th2}),
(\ref{GS21}) and (\ref{cotanh2}).

To analyze solutions for exited states we rewrite superpotential
(\ref{tan}) in terms of variable $y$:
\begin{gather}\label{tan3}W_{\kappa,\mu}=\lambda\left(\kappa(2y-1)+
\frac\mu2\sqrt{y(1-y)}\so+\frac\omega\kappa\st\right).\end{gather}
 We see that $W_\kappa$ is nonsingular at $y=0$ and $y=1$, the same is true for
 $W_{\kappa+n,\mu}$ for any natural number $n$. Functions (\ref{GS3})
 are still regular at these points if we change $\kappa\to\kappa+n$,
 thus we can again apply relations (\ref{exit1})--(\ref{psinn}) to
 prove the normalizability of the corresponding solutions (\ref{psin})
 for arbitrary $n$.

In a similar way we can prove the square integrability for solutions
(\ref{psin}) corresponding to ground state solutions (\ref{tan2}),
(\ref{GS3d}), (\ref{tan2d}), (\ref{GS21}), (\ref{cotanh2}),
(\ref{GS2}) and (\ref{th2}). However in these cases we again have
constraint  (\ref{th1}) which generates restriction (\ref{nko}) for
the number $n$ enumerating the exited states (\ref{psin}).
Analogously, starting with condition (\ref{condmu}) we come to
conclusion that solutions (\ref{psinka}) for the Schr\"odinger
equation with potential (\ref{pot5})  are square integrable provided
quantum number $n$ satisfies the condition
\begin{gather}n<|\mu|-\sqrt{\omega}-\frac12\label{nmo}.\end{gather}

Thus for any fixed $\kappa$, $\mu$ and $\omega$ equation (\ref{se2})
with potentials (\ref{pot2}) (\ref{pot5}) and (\ref{pot4}) describes
a system which has a finite number of states with discrete spectrum.
These states are enumerated by non-negative  natural numbers $n$
satisfying condition (\ref{nko}) or (\ref{nmo}).

The systems with potentials (\ref{pot1})--(\ref{pot4}) can also have
states with continuous spectrum. In particular, such states should
change the  bound states when conditions (\ref{nko}) and (\ref{nmo})
are violated. Analysis of the states with
  continuous spectra lies out of frames of the present paper.

\section{Discussion\label{disscus}}

Generalizing the supersymmetric PS problem we find a family of
matrix potentials for Shr\"odinger equation satisfying the shape
invariance condition. In this way we find five  exactly solvable
problems for systems of coupled Shr\"odinger equations. The related
matrix potentials are given by equations (\ref{pot1})--
(\ref{pot4}).

 Let us
stress that we present the completed classification of shape
invariant superpotentials of the generic form (\ref{SP}) where $P$
and $R$ are hermitian matrices of arbitrary finite dimension and $Q$
is proportional to the unit matrix. Namely, we show that such
objects can be reduced to direct sums of known scalar
superpotentials and superpotentials presented in section \ref{SIP}.

The found potentials include  parameters $\lambda, \kappa, \mu$ and
$\omega$ whose possible values are restricted but quite arbitrary.
Moreover, parameters $\omega$ in (\ref{pot1}) and $\mu$ in
(\ref{pot2}) can be reduced to unity by scaling and shifting the
independent variable $x$ correspondingly.

Taking into account all possibilities enumerated in (\ref{condk1}),
(\ref{SS5}), (\ref{SS55}) and (\ref{condk4})--(\ref{condk6})
 we conclude that in the case of potentials (\ref{pot1}), (\ref{pot3})
 and (\ref{pot5}) there are discrete spectrum states   for
  all real
values of arbitrary parameters $\lambda, \kappa$ and  $\mu $
 except the case $\kappa=\mu$.  Parameter $\omega$ can be constrained
 by equations (\ref{th1}) or (\ref{condmu}).

Potential (\ref{pot1}) is a slightly generalized effective potential
for the PS problem. Moreover, these potentials coincide for a
particular value $\mu=0$ of arbitrary parameter $\mu$.
However, if $\mu\neq0$  potential (\ref{pot1}) is not
equivalent to the potential appearing in the PS problem and
corresponds to a more general interaction in the initial
three-dimension problem.

 At the best of our
knowledge the remaining potentials (\ref{pot2})--(\ref{pot4}) are
new. The related Schr\"odinger equations can be integrated using
tools of the SUSY quantum mechanics. The corresponding spectrum and
eigenvectors are given by equations (\ref{EV0})--(\ref{Nnm}) and
(\ref{psin}) or (\ref{psinka}) while the ground state solutions are
discussed in section \ref{GSS}. solutions Notice that the "matrix
supersymmetry" has a new feature in comparison with the standard
(i.e., scalar) one. Namely, matrix models with shape invariance can
have degenerated ground states  in spite of that there exists a
normalizable solution for equation (\ref{psi0}). Example of  models
with such specific spontaneously broken SUSY is given by the
Schr\"odinger equation (\ref{se2}) with potentials
(\ref{pot3})--(\ref{pot4}).

Mathematically, there are natural reasons for appearance of a zero
energy doublet of the ground states in systems with the matrix
supersymmetry. The thing is that equation (\ref{psi0}) is a system
of {\it two} the first order equations whose solutions are linear
combinations of {\it two} functions while in the ordinary SUSY
quantum mechanics we have a one first order equation for ground
states. For potentials (\ref{pot1}) and (\ref{pot2}) only one of
these functions is normalizable but for potentials
(\ref{pot3})--(\ref{pot2}) there are two ground state solutions.

Let us note that existence of zero energy doublets of the ground
states was already registered in periodic quantum systems, see
\cite{ground_degeneration} and \cite{plu1} for discussion of this
phenomenon. In this connection it seems to be interesting to extend
our approach to the case of periodic systems. Formally speaking, the
only new constructive elements of potentials
(\ref{pot1})--(\ref{pot4}) in comparison with the standard scalar
shape invariant potentials are matrices $\so$ and $\st$
which are involutions anticommuting between themselves. In fact the
nature of these involutions is not essential for deducing the shape
invariant potentials, and many of the results discussed in present
paper can be generalized to the case of  another involutions. For
example, it is possible to change the mentioned matrices by
reflection and shift operators which also can be anticommuting
involutions being applied to functions with an appropriate parity
and periodicity. In this way it seems to be possible to extend the
list of  potentials which admit supersymmetry including shifts of
arguments \cite{plu2}. A classification of anticommuting  discrete
symmetries and the corresponding supersymmetric versions of the
Schr\"odinger and Pauli equations can be found in
\cite{nik}.

An interesting phenomena which appears to be  typical for systems
with matrix SUSY is the dual shape invariance discussed in Section
\ref{dsi}. It enables to impose much less restrictive constrains on
parameters of potentials then the ordinary shape invariance. In
addition, it can be used to explain the insensibility of the spectra
(\ref{EV0})--(\ref{EV0}), (\ref{Nn}) on parameter $\mu$. Namely,
hamiltonians with shifted $\mu$ should be almost isospectral thanks
to the dual shape invariance, which is incompatible with
$\mu$-dependence of energy values (excluding the exotic case when
these values are periodic functions of $\mu$).

For some values of parameters $\mu$ and $\kappa$  the additional
branch of spectrum caused by the dual shape invariance can appear.
In particular it is true for potential (\ref{pot1}) with $\mu=0$ and
$0<\kappa<1/2$. Enhanced analysis of such potentials was made in
paper \cite{Ioffe}. In the present paper we slightly refine results
of \cite{Ioffe}.

Let us note that the dual shape invariance can be recognized for two
potentials of the ordinary SUSY quantum mechanics, namely, for the
trigonometric Scraft 1 and generalized P\"oshl-Teller potentials:
\begin{gather*}V_1=(\kappa(\kappa-1)+\mu^2)\sec^2
x+\mu(1-2\kappa)\sec\lambda x \tan\lambda
x,\\V_2=(\kappa(\kappa-1)+\mu^2) \csch^2(
x)+\mu(1-2\kappa)\coth\lambda x\csch x
\end{gather*}
both of which admit symmetries (\ref{change}). The corresponding
energy spectra is $\mu$-independent also, and the dual shape
invariance can be used to explain this phenomena.

It is shown in section \ref{iso} that for some values of parameters
$\mu$ and $\kappa$ the matrix potentials (\ref{pot1})--(\ref{pot4})
are isospectral with direct sums of one dimensional shape invariant
potentials. Unfortunately, in this way we cannot establish
isospectrality with the reflectionless hyperbolic Poschl-Teller
(HPT) system  which has a lot of interesting applications and admits
a hidden (bosonized) nonlinear supersymmetry \cite{plu3}. A natural
question arises if there are other matrix superpotentials which can
add the list given by equations (\ref{inverse})--(\ref{cotanh}) and
may include a matrix counterpart of the reflectionless HPT
potential?

In sections 3--5 we show that irreducible matrix potentials of
generic form (\ref{SP}) are exhausted by known scalar ones and
$2\times2$ matrix operators given by equations
(\ref{inverse})--(\ref{cotanh}). Of course we restrict ourselves to
the shape invariance of type 1 when variable parameter $\kappa$ is
changing by shifts.

 Nevertheless it is possible to search for
matrix superpotentials in a more general approach, when matrix $Q$
in (\ref{SP}) is not restricted to be proportional to the unit one.
Let us present an example of such superpotential:
\[W_{\kappa,\mu}=\left(\kappa+\frac12\right)\frac{
x-c\st}{c^2-x^2}+\frac\omega{(2\kappa+1)}\so\] where $c$
is a a constant. The corresponding potential $V_\kappa$
(\ref{pfactorization}) is shape invariant and has the following
form:
\[V_\kappa=(4\kappa^2-1)\frac{x^2+c^2-2cx\st}{4(x^2-c^2)^2}+\frac{\omega
x}{c^2-x^2}\so\] while the energy spectrum is given by equation
(\ref{EV0}). The other examples (which, however, are restricted to a
linear dependence of $W_{\kappa,\mu}$ on variable parameter
$\kappa$) can be found in paper \cite{Fu}.

 The problem of classification of matrix potentials (\ref{SP}) with generic
 hermitian matrices $Q, P$ and $R$ is a subject of our contemporary research.

\end{document}